# Perovskite-perovskite tandem photovoltaics with optimized bandgaps

Giles E. Eperon[1,3],* Tomas Leijtens[2],* Kevin A. Bush[2], Rohit Prasanna,[2] Thomas Green[1], Jacob Tse-Wei Wang[1], David P. McMeekin[1], George Volonakis[4], Rebecca L. Milot[1], Richard May[2], Axel Palmstrom[2], Daniel J. Slotcavage[2], Rebecca A. Belisle[2], Jay B. Patel[1], Elizabeth S. Parrott[1], Rebecca J. Sutton[1], Wen Ma,[5] Farhad Moghadam,[5] Bert Conings[1,6], Aslihan Babayigit[1,6], Hans-Gerd Boyen[6], Stacey Bent[2], Feliciano Giustino[4], Laura M. Herz[1], Michael B. Johnston[1], Michael D. McGehee[2] and Henry J. Snaith[1]

*GEE and TL contributed equally to this work.

Correspondence to: MDM (mmcgehee@stanford.edu), HJS (henry.snaith@physics.ox.ac.uk)

[1]Department of Physics, University of Oxford, Clarendon Laboratory, Parks Road, Oxford OX1 3PU, UK
[2]Department of Materials Science, Stanford University, Lomita Mall, Stanford, CA, USA
[3]Department of Chemistry, University of Washington, Seattle, WA, USA
[4]Department of Materials, University of Oxford, Parks Road, Oxford OX1 3PH, UK
[5]SunPreme, Palomar Avenue, Sunnyvale, CA, USA
[6]Institute for Materials Research, Hasselt University, Diepenbeek, Belgium

**We demonstrate four and two-terminal perovskite-perovskite tandem solar cells with ideally matched bandgaps. We develop an infrared absorbing 1.2eV bandgap perovskite, $FA_{0.75}Cs_{0.25}Sn_{0.5}Pb_{0.5}I_3$, that can deliver 14.8 % efficiency. By combining this material with a wider bandgap $FA_{0.83}Cs_{0.17}Pb(I_{0.5}Br_{0.5})_3$ material, we reach monolithic two terminal tandem efficiencies of 17.0 % with over 1.65 volts open-circuit voltage. We also make mechanically stacked four terminal tandem cells and obtain 20.3 % efficiency. Crucially, we find that our infrared absorbing perovskite cells exhibit excellent thermal and atmospheric stability, unprecedented for Sn based perovskites. This device architecture and materials set will enable "all perovskite" thin film solar cells to reach the highest efficiencies in the long term at the lowest costs.**

Metal halide perovskites ($ABX_3$, where A is typically Cs, methylammonium (MA), or formamidinium (FA), B is Pb or Sn, and X is I, Br, or Cl) have emerged as an extremely promising photovoltaic (PV) technology due to their rapidly increasing power conversion efficiencies (PCEs) and low processing costs. Single junction perovskite devices have reached a certified 22% PCE(*1*), but the first commercial iterations of perovskite PVs will likely be as an "add-on" to silicon (Si) PV. In a tandem configuration, a perovskite with a band gap of ~1.75 eV can enhance the efficiency of the silicon cell.(*2*) An all-perovskite tandem cell could deliver lower fabrication costs, but requires band gaps that have not yet been realized. The highest efficiency tandem devices would require a rear cell with a band gap of 0.9 to 1.2 eV and a front cell with a band gap of 1.7 to 1.9 eV. Although materials such as $FA_{0.83}Cs_{0.17}Pb(I_xBr_{1-x})_3$ deliver appropriate band gaps for the front cell (*2*), Pb-based materials cannot be tuned to below 1.48 eV for the rear cell. Completely replacing Pb with Sn can shift the band gap to ~1.3eV (for $MASnI_3$)(*3*), but the tin-based materials are notoriously air-sensitive and difficult to process, and PV devices based on them have been limited to ~6% PCE.(*3*, *4*) An anomalous band gap bowing in mixed tin-lead perovskite systems ($MAPb_{0.5}Sn_{0.5}I_3$) has given band gaps of ~1.2eV but mediocre performance (~7% PCE). Very recently, PCE of over 14% has been reported with $MA_{0.5}FA_{0.5}Pb_{0.75}Sn_{0.25}I_3$ cells, for band gaps > 1.3 eV and all-perovskite 4 terminal tandem cells with 19% efficiency.(*5*)(*6*)(*7*). Here, we demonstrate a stable 14.8 % efficient perovskite solar cell based on a 1.2 eV bandgap $FA_{0.75}Cs_{0.25}Pb_{0.5}Sn_{0.5}I_3$ absorber. We measure open-circuit voltages ($V_{OC}$s) of up to 0.83 V in these cells, which represents a smaller voltage deficit between band gap and $V_{OC}$ than measured for the highest efficiency lead based perovskite cells. We then combined these with 1.8 eV $FA_{0.83}Cs_{0.17}Pb(I_{0.5}Br_{0.5})_3$ perovskite cells, to demonstrate current matched and efficient (17.0 %) monolithic all-perovskite 2-terminal tandem solar cells on small areas and 13.8 % on large areas, with $V_{OC}$ > 1.65 V. Finally, we fabricated 20.3 % efficient small area and 16.0 % efficient 1cm$^2$ all-perovskite four-terminal tandems using a semitransparent 1.6eV $FA_{0.83}Cs_{0.17}Pb(I_{0.83}Br_{0.17})_3$ front cell.

**Results and discussion**

It has proven difficult to fabricate smooth, pinhole-free layers of tin-based perovskites on planar substrates (Fig S1).(*3*, *4*) We developed a technique, precursor-phase antisolvent immersion (PAI), to deposit uniform layers of tin-containing perovskites, $FASn_xPb_{1-x}I_3$, that combines two previous methods: the use of low vapor pressure solvents to retard crystallization by forming precursor complexes and an anti-solvent bath to crystallize the film with only gentle heating.(*4*, *8*) Rather than using neat dimethyl sulfoxide (DMSO) as a solvent (*4*), a mixture of DMSO and dimethylformamide (DMF) allowed spin-coating of a uniform transparent precursor film that was not yet fully crystallized. Immersion of the films immediately in an antisolvent bath (anisole) (Fig S2) rapidly changed the film to a deep red.(*9*) Subsequent annealing at 70 °C removed residual DMSO (Fig S7) to form smooth, dark, highly crystalline and uniform $FASn_xPb_{1-x}I_3$ films over the entire range of values of x = 0 to 1 (Fig. 1A). Only the neat Pb perovskite required heating at a higher temperature (170 °C) to convert the film from the yellow room-temperature phase to the black phase.(*10*)

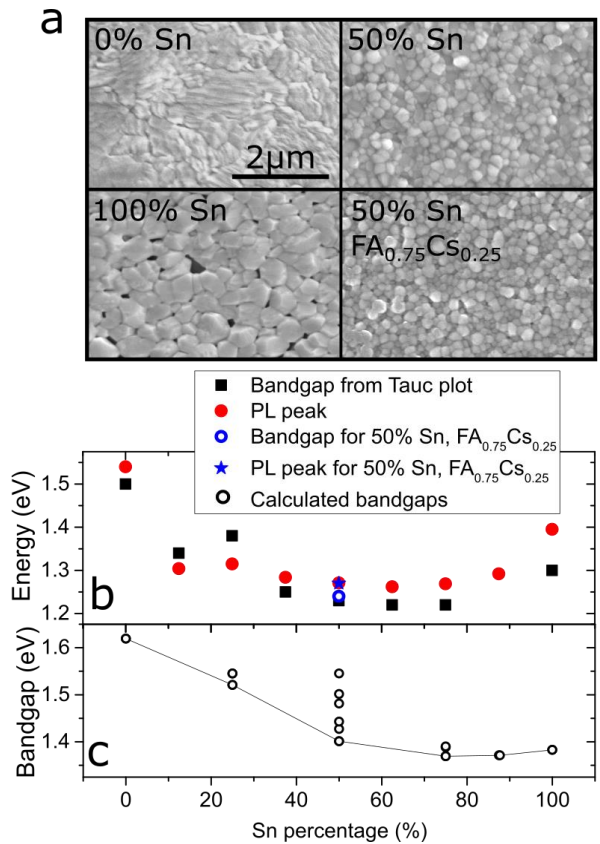

*Figure 1: Tin-lead alloying a) Scanning electron microscope (SEM) images showing the top surface of $FASn_xPb_{1-x}I_3$ films with different Sn percentages and $FA_{0.75}Cs_{0.25}Pb_{0.5}Sn_{0.5}I_3$ (discussed later, and labelled here as "50%, $FA_{0.75}Cs_{0.25}$"), fabricated with the PAI deposition technique. The 0% Sn films were annealed at 170 ºC, while the other films were heated at 70 ºC. b) Plot of experimentally estimated bandgap as a function of Sn %, determined from absorption onset in a Tauc plot (assuming direct bandgap) of the absorption (black); PL peak positions are given in red. c) Bandgaps for Sn-Pb perovskite alloys calculated from first principles using a supercell containing eight $BX_6$ octahedra, where the Sn and Pb atoms are ordered relative to each other (See SM for full details). Points plotted represent all possible bandgaps for a particular composition, based on all possible Sn-Pb configurations; a solid line is drawn through the lowest bandgap options as a comparison to experiment.*

Photoluminescence (PL) spectra and absorption spectra of a range of compositions (Fig. S4) allowed us to estimate the optical band gap from Tauc plots (from absorption) and the PL peak positions (Fig. 1B).(*9*) The band gap narrowed between the two composition endpoints, similar to the observations of Kanatzidis et al. with the MA system (*5*), and between 50 to 75% Sn, was almost 1.2 eV. X-ray diffraction (XRD) spectra (Fig S5) for the whole series revealed a single dominant perovskite phase (see Table S1).

To understand this anomalous band gap trend, we performed first-principles calculations of band gaps as a function of the tin-lead ratio (details in the SM).(*9*) For a disordered solid solution with Pb and Sn in random locations, the calculated band gap decreased monotonically (Fig. S8). For an ordered structure, we placed the Sn and Pb atoms in specific positions relative to each other within a repeating lattice unit of eight octahedra in a 'supercell'. Here, if we took the lowest band gaps for each ratio, an anomalous bandgap trend emerges (Fig. 1C). We elucidate that for compositions with more than > 50% Sn, a specific type of short-range order in the Pb-Sn positions allowed the band gap to dip below the end points. Im et al. attributed a similar band gap trend observed for $MAPb_xSn_{1-x}I_3$ to the competition between spin-orbit coupling and distortions of the lattice (*11*), but if this was the case here, we should have observed it in the random solid solution approach. The energetic difference between the various Pb-Sn configurations was on the order of 2 meV, so at room temperature the materials are likely to contain various combinations of the configurations which we show in Fig. S7, but the absorption and emission onsets reflect the regions with the smallest gap.

In order to determine the diffusion length, mobility and recombination lifetimes of these materials, we performed optical pump-probe terahertz (THz spectroscopy) on $FASnI_3$ and $FASn_{0.5}Pb_{0.5}I_3$. The fluence dependence of the THz transients for $FASn_{0.5}Pb_{0.5}I_3$ (Fig. S9) exhibited faster decays at higher intensities as the result of increased bimolecular and auger recombination.(*12*) We calculated the recombination rate constants and the charge-carrier mobilities, which were 22 and 17 $cm^2V^{-1}s^{-1}$ for $FASnI_3$ and $FASn_{0.5}Pb_{0.5}I_3$, respectively, comparable to values for Pb perovskite films (*12*, *13*). In comparison, for $MASnI_3$, the value was only 2 $cm^2V^{-1}s^{-1}$(*3*, *14*). For charge-carrier densities typical under solar illumination, charge-carrier diffusion lengths of ~300nm were reached for $FASn_{0.5}Pb_{0.5}I_3$(details in the SM). Although lower than that for the best reported perovskite materials, it is equivalent to the typical thickness required to absorb most incident light (~300 to 400nm)(*15*).

We fabricated a series of planar heterojunction devices in the 'inverted' p-i-n architecture (*16*) comprising ITO/Poly(3,4-ethylenedioxythiophene)-poly(styrenesulfonate) (PEDOT:PSS)/$FASn_xPb_{1-x}I_3$/$C_{60}$/bathcuproine (BCP) capped with an Ag or Au electrode (Fig. 2A). The current-density voltage (J-V) curves and external quantum efficiency (EQE) measurements for the whole compositional series are shown in Fig. S10. The onset of the EQEs closely matched that of the absorption of the materials, with light harvested out to ~1020nm in the 50%-75% Sn compositions. The highest efficiencies are generated from the devices with 50% Sn, within the lowest band gap region, thus we used this material to optimize our low-gap solar cells.

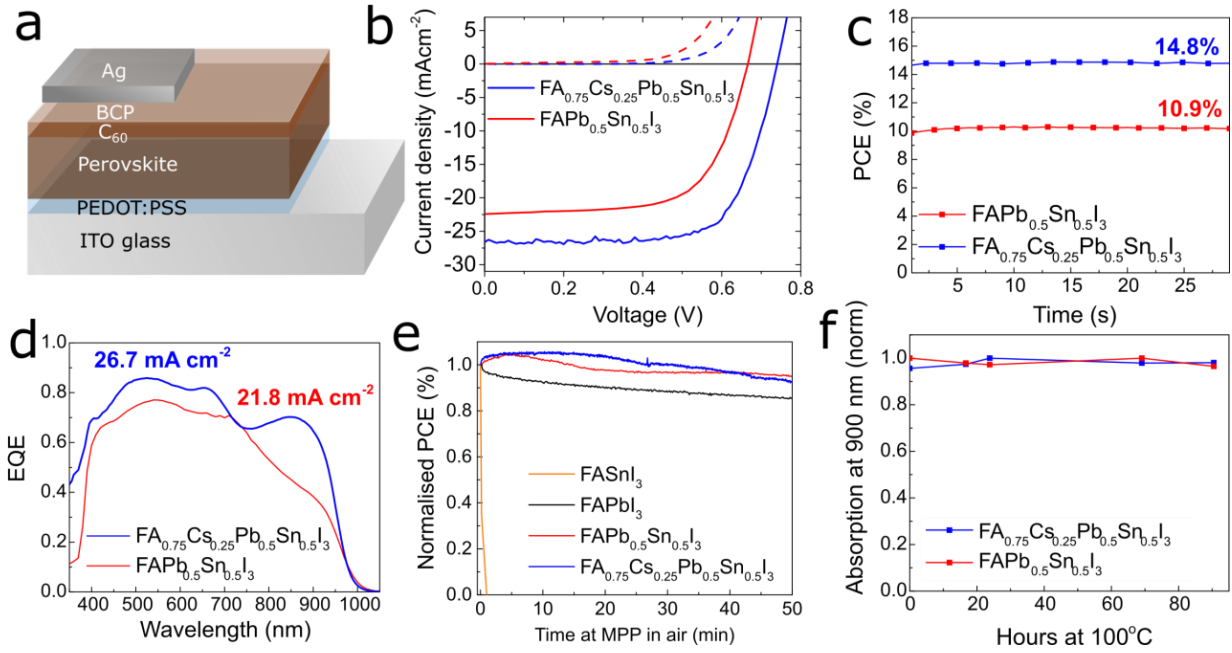

*Fig. 2. Performance and stability of FASn$_{0.5}$Pb$_{0.5}$I$_3$ and FA$_{0.75}$Cs$_{0.25}$Sn$_{0.5}$Pb$_{0.5}$I$_3$ perovskite solar cells. a) Schematic of the device architecture for narrow-gap single junction perovskite solar cells. b) Current-voltage characteristics under AM1.5G illumination for the champion FASn$_{0.5}$Pb$_{0.5}$I$_3$ and FA$_{0.75}$Cs$_{0.25}$Sn$_{0.5}$Pb$_{0.5}$I$_3$ devices under illumination (solid lines) and in the dark (dotted lines), measured at 0.1V/s with no pre-biasing or light soaking. c) Champion solar cells stabilized power output, measured via a maximum power point tracking algorithm. d) External quantum efficiency for the champion devices of each material with the integrated current shown as an inset, providing a good match to the JV scan J$_{sc}$. e) PCE as a function of time for three compositions of FASn$_x$Pb$_{1-x}$I$_3$ (x=0, 0.5, 1) as well as FA$_{0.75}$Cs$_{0.25}$Sn$_{0.5}$Pb$_{0.5}$I$_3$ measured by holding the cell at maximum power point in air under AM1.5 illumination. f) Thermal stability of FASn$_{0.5}$Pb$_{0.5}$I$_3$ and FA$_{0.75}$Cs$_{0.25}$Sn$_{0.5}$Pb$_{0.5}$I$_3$ films, quantified by heating the samples at 100 °C and monitoring their absorption at 900 nm as a function of time.*

|  | J$_{SC}$ (mA cm$^{-2}$) | V$_{OC}$ (V) | FF | PCE (%) | SPO (%) |
|---|---|---|---|---|---|
| **FASn$_{0.5}$Pb$_{0.5}$I$_3$** | 21.9 | 0.70 | 0.66 | 10.2 | **10.9** |
| **FA$_{0.75}$Cs$_{0.25}$Sn$_{0.5}$Pb$_{0.5}$I$_3$** | 26.7 | 0.74 | 0.71 | 14.1 | **14.8** |

*Table 1. Device parameters corresponding to the J-V curves in Figure 2b.*

A small addition of Cs boosts the performance and stability of Pb based perovskites.(*17–19*) Substituting 25% of the FA with Cs in our films had little impact on band gap, morphology, PL, crystal structure, and charge carrier diffusion lengths (Fig. 1, Fig. S4, and Fig. S7), but device performance was enhanced (Fig. 2, B to D). The best FASn$_{0.5}$Pb$_{0.5}$I$_3$ device yields 10.9% PCE, whereas the best FA$_{0.75}$Cs$_{0.25}$Sn$_{0.5}$Pb$_{0.5}$I$_3$ device exhibited an impressive short-circuit current of 26.7 mA cm$^{-2}$, 0.74 V V$_{OC}$, and 0.71 FF to yield 14.1 % PCE. We note that the processing of each composition was optimized separately. These devices did not exhibit appreciable rate-dependent hysteresis and the stabilized power output (14.8%) matched the scanned performance well. This efficiency was comparable with the very best solution-processed low band gap copper-indium-gallium-diselenide (CIGS) solar cells.(*20*) Thus, FA$_{0.75}$Cs$_{0.25}$Pb$_{0.5}$Sn$_{0.5}$I$_3$ is well suited for a rear junction in a solution processed tandem solar cell, without the need for high temperature thermal processing.

We performed UPS and XPS measurements to determine the energetic positions of the conduction and valence bands (Fig S11). The band levels for FASn$_{0.5}$Pb$_{0.5}$I$_3$ are well matched for C60 and PEDOT:PSS as electron and hole acceptors. The Cs-containing material showed an energetically shallower valence band and mild p-type doping (*21, 22*).

The electronic losses in a solar cell are reflected by the difference in energy between the band gap of the absorber and V$_{oc}$ (the loss-in-potential) (*23*) . For crystalline silicon PV cells, which generate a record V$_{oc}$ of 0.74 V and have a bandgap of 1.12 eV, this loss is 0.38 V.(*24*) Some of our FA$_{0.75}$Cs$_{0.25}$Sn$_{0.5}$Pb$_{0.5}$I$_3$ devices here, with a thinner active layer, displayed V$_{oc}$s up to 0.833 V (Fig. S11), with a 1.22 eV bandgap, exhibiting a comparable loss in potential of 0.386 eV.

Tin based perovskites have previously been observed to be extremely unstable in air (*25*), so we carried out a simple aging test on the FASn$_{0.5}$Pb$_{0.5}$I$_3$ and FA$_{0.75}$Cs$_{0.25}$Pb$_{0.5}$Sn$_{0.5}$I$_3$ devices. We held the devices at maximum power point under 100 mWcm$^{-2}$ illumination and measured power output over time in ambient air with a relative humidity of 50±5% (Fig. 2E). The FAPbI$_3$ device maintains its performance relatively well with a small drop observed over the time (to 85% of initial PCE over 50 minutes), possibly associated with photo-oxidation and hydration of the un-encapsulated perovskite layer, or a partial reversion to the yellow room-temperature phase.(*17*, *26*) Both the FASn$_{0.5}$Pb$_{0.5}$I$_3$ and FA$_{0.75}$Cs$_{0.25}$Sn$_{0.5}$Pb$_{0.5}$I$_3$ showed similar or even better stability than the neat Pb material. We also subjected bare perovskite films to thermal stress, heating for 4 days at 100 °C under nitrogen; there were no changes in absorption spectra, a monitor of optical quality and hence stability(Fig. 2F) (*27*). We also monitored the performance of full devices at 85°C over several months (Figure S13), and found that the Sn:Pb material displays similar device stability as the neat Pb material. The contribution of both Sn and Pb orbitals to the valance band minimum may reduce the propensity of Sn$^{2+}$ to oxidize to Sn$^{4+}$.

A 1.2eV perovskite is ideally suited as the rear cell in either monolithic two terminal (2T) tandem solar cells or mechanically stacked four terminal (4T) tandem solar cells (Fig. 3A). The subcells in a 2T tandem must be current matched to deliver optimum performance, and connected with a recombination layer. A 4T tandem operates the two cells independently, but requires an extra transparent electrode which can result in more absorption losses and higher cost. Theoretical efficiencies using a 1.2 eV rear cell (Figure S14) show that the 2T architecture requires the use of a top cell with a ~1.75-1.85 eV bandgap whereas the 4T architecture has a much more relaxed requirement of 1.6 to 1.9 eV.

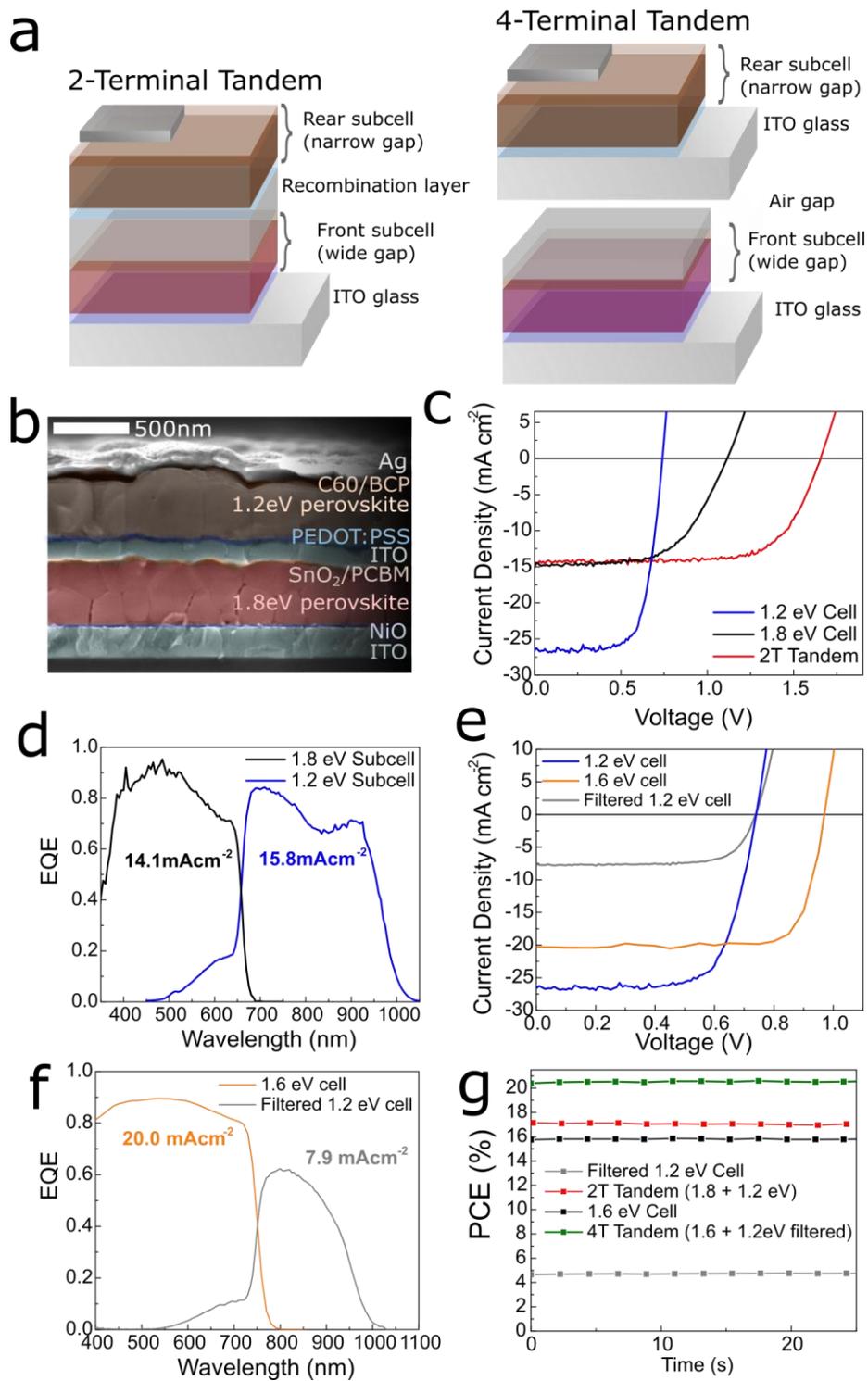

*Figure 3. Perovskite-perovskite tandems.* a) Schematics showing 2- and 4-terminal tandem perovskite solar cell concepts. In this image devices would be illuminated from below. b) Scanning electron micrograph of the two-terminal perovskite-perovskite tandem. c) Scanned current-voltage characteristics under AM 1.5G illumination, of the two-terminal perovskite-perovskite tandem, of the 1.2 eV solar cell, and the ITO capped 1.8 eV solar cell. d) External quantum efficiency spectra for the sub-cells. e) J-V curves of a 1.2 eV perovskite, of that same solar cell filtered by an ITO capped 1.6 eV perovskite solar cell, and the ITO capped 1.6 eV perovskite solar cell, used to determine the mechanically stacked tandem efficiency. f) External quantum efficiency spectra for the mechanically stacked tandem. g) The stabilized power output tracked over time at maximum power point for the 2T perovskite solar cell, the 1.2 eV perovskite solar cell filtered by an ITO capped 1.6 eV perovskite solar cell, the ITO capped 1.6 eV perovskite solar cell, and the mechanically stacked tandem under AM 1.5G illumination. The SPO for the 1.8 eV subcell is plotted in Figure S14 and given in Table 2.

|  | $J_{SC}$ (mA cm$^{-2}$) | $V_{OC}$ (V) | FF | PCE (%) | SPO (%) |
|---|---|---|---|---|---|
| **1.2 eV cell** | 26.7 | 0.74 | 0.71 | 14.1 | **14.8** |
| **1.8 eV cell** | 15.1 | 1.12 | 0.58 | 9.8 | **9.5** |
| **2T tandem** | 14.5 | 1.66 | 0.70 | 16.9 | **17.0** |
|  |  |  |  |  |  |
| **Filtered 1.2 eV cell** | 7.9 | 0.74 | 0.73 | 4.4 | **4.5** |
| **ITO capped 1.6 eV cell** | 20.3 | 0.97 | 0.79 | 15.7 | **15.8** |
| **4T tandem** | - | - | - | 20.1 | **20.3** |
|  |  |  |  |  |  |
| **1 cm$^2$ 2T tandem** | 13.5 | 1.76 | 0.56 | 13.3 | **13.8** |
| **1 cm$^2$ 4T tandem** | - | - | - | 16.4 | **16.0** |

*Table 2*. Solar cell performance parameters corresponding to the J-V curves shown in Figure 3. Cell active areas are 0.20 or 1 cm$^2$. SPO = stabilized power output from MPP tracking. Large area tandem data are plotted in Figure S19.

We can obtain efficient and stable perovskites with appropriate wide band gaps for front cells in tandem architectures by using a mixture of FA and Cs cations (*2*) and control the band gap by tuning the Br:I ratio; FA$_{0.83}$Cs$_{0.17}$Pb(I$_{0.5}$Br$_{0.5}$)$_3$ has a 1.8 eV band gap ideally suited for the 2T tandem. However, their higher losses in potential than the more commonly used 1.6 eV perovskites (*28*) make the latter more suited for the 4T tandem. We prepared both of these perovskites in the p-i-n structure depicted in Figure 3A, using NiO$_x$ and phenyl-C61-butyric acid methyl ester (PCBM) as the hole and electron contacts respectively. We applied the PAI deposition route to form smooth and thick perovskite layers, obtaining efficient devices with appropriate photocurrents and voltages up to 1.1V (see Fig. S14).(*9*)

For the recombination layer in the 2T cell, we used a layer of tin oxide coated with sputter coated indium tin oxide (ITO)(*29*). This ITO layer completely protects the underlying perovskite solar cell from any solvent damage (Fig S16), meaning we could fabricate the 1.2 eV FA$_{0.75}$Cs$_{0.25}$Sn$_{0.5}$Pb$_{0.5}$I$_3$ solar cell directly on top. We plot the JV curves of the best single junction 1.2 eV cells, single junction 1.8 eV cell, and that of the best 2T tandem device in Fig. 3C. We observed good performance for the 2T tandem solar cells, in excess of either of the individual subcells and impressive considering the somewhat non-optimized 1.8eV top cell. The photocurrent of the tandem solar cell was 14.5mAcm$^{-2}$, voltage is an appropriate addition of the two subcells (1.66 V), and the fill factor is 0.70, yielding an overall performance of 16.9 % via a scanned JV curve and of 17.0 % when stabilized at its maximum power point. None of the devices exhibit substantial hysteresis in the JV curves (Fig. 3G and Figure S17).

The photocurrent is remarkably high when compared to the photocurrent density of the best reported monolithic perovskite-silicon tandems.(*30*, *31*) EQE measurements (Fig. 3D) demonstrate that the two subcells are fairly well matched, with the wide gap subcell limiting the current. One benefit of a tandem architecture, that we observe here, is that the FF tends not to be limited to the lowest value of the individual subcells, due to the reduced impact of series resistance on a higher-voltage cell.(*32*) Furthermore, we held a 2T tandem at its maximum power point under illumination in nitrogen for more than 18 hours and it showed effectively no performance drop (Fig. S18).

For a 4T tandem, we used an efficient 1.6eV band gap FA$_{0.83}$Cs$_{0.17}$Pb(I$_{0.83}$Br$_{0.17}$)$_3$ perovskite, similar to that reported by McMeekin et al but in p-i-n configuration (*2*) with a transparent ITO top contact. We obtained a 15.8 % efficient solar cell with a V$_{oc}$ ~1 V, and when we use it to filter a 14.8 % FA$_{0.75}$Cs$_{0.25}$Sn$_{0.5}$Pb$_{0.5}$I$_3$ cell, we can still extract substantial photocurrent (7.9 mA cm$^{-2}$) from the low-bandgap device. We plot the JV curves and EQE spectra of the 1.6 and 1.2 eV cells in the 4T tandem in Fig 3E and F, and show we can obtain an additional 4.5 % PCE from the 1.2 eV rear cell, yielding an overall stabilized tandem efficiency of 20.3 % (Fig. 3G).

The above results were for 0.2cm$^2$ devices. We also made large-area (1cm$^2$) versions of the single junctions, 2T and 4T tandems and show the current-voltage characteristics in Table 2 and Fig. S18, with 2T tandem 1cm$^2$ devices exhibiting 13.8 % stabilized PCE and 4T tandems 16.0 %. The 17.0 % PCE 2T and 20.3 % 4T tandems, which are for devices that could be further optimized, are already far in excess of the best tandem solar cells made with other similarly low cost semiconductors, such as those made with organic small molecules (world record 13 %) or amorphous and microcrystalline silicon (13.5%).(*1*, *24*) Notably, our results illustrate that the tandem cell should be at least 4 to 5% more efficient than the best 1.6 eV single junction perovskite cells, indicating that as the efficiency of the single junction cells increases, then the tandem approach will enable this low temperature-processed polycrystalline thin film technology to surpass the 30% efficiency barrier.

**Acknowledgments**

We thank M. T. Hörantner for performing the Shockley-Queisser calculation. The research leading to these results has received funding from the Graphene Flagship (EU FP7 grant no. 604391), the Leverhulme Trust (Grant RL-2012-001), the U.K. Engineering and Physical Sciences Research Council (Grant No. EP/ J009857/1 and EP/M020517/1), and the European Union Seventh Framework Programme (FP7/2007-2013) under grant agreement nos. 239578 (ALIGN) and 604032 (MESO). TL is funded by a Marie Sklodowska Curie International Fellowship under grant agreement H2O2IF-GA-2015-659225. AB is financed by IMEC (Leuven) in the framework of a joint PhD programme with Hasselt University. BC is a postdoctoral research fellow of the Research Fund Flanders (FWO). We also acknowledge the Office of Naval Research USA for support. We acknowledge the use of the University of Oxford Advanced Research Computing (ARC) facility (http://dx.doi.org/10.5281/zenodo.22558) and the ARCHER UK National Supercomputing Service under the "AMSEC" Leadership project. We thank the Global Climate and Energy Project (GCEP) at Stanford University. All data pertaining to the conclusions of this work is present in the main paper and the supplementary material.